# Teaching Fluid Mechanics for Undergraduate Students in Applied Industrial Biology: from Theory to Atypical Experiments

Rafik ABSI [1], Caroline NALPAS [1], Florence DUFOUR [1], Denis HUET [1], Rachid BENNACER [2], Tahar ABSI [3]

[1] *EBI - Ecole de Biologie Industrielle, Inst. Polytech. St-Louis, 32 Bd du Port, 95094 Cergy-Pontoise, France.*
E-mail : r.absi@ebi-edu.com
[2] *LMT-ENS Cachan, Ecole Normale Supérieure, 61 av. du président Wilson F-94235 Cachan Cedex, France.*
[3] *Département de Psychologie et des Sciences de l'Education, Université d'Alger, Algérie.*

**Abstract**

EBI is a further education establishment which provides education in applied industrial biology at level of MSc engineering degree. Fluid mechanics at EBI was considered by students as difficult who seemed somewhat unmotivated. In order to motivate them, we applied a new play-based pedagogy. Students were asked to draw inspiration from everyday life situations to find applications of fluid mechanics and to do experiments to verify and validate some theoretical results obtained in course. In this paper, we present an innovative teaching/learning pedagogy which includes the concept of learning through play and its implications in fluid mechanics for engineering. Examples of atypical experiments in fluid mechanics made by students are presented. Based on teaching evaluation by students, it is possible to know how students feel the course. The effectiveness of this approach to motivate students is presented through an analysis of students' teaching assessment. Learning through play proved a great success in fluid mechanics where course evaluations increased substantially. Fluid mechanics has been progressively perceived as interesting, useful, pleasant and easy to assimilate. It is shown that this pedagogy which includes educational gaming presents benefits for students. These experiments seem therefore to be a very effective tool for improving teaching/learning activities in higher education.

1.   **Introduction**

Teaching science and technology has been often related to experiments conducted to confirm or disprove a theory. If the path travelled by the scientists is marked by different experiments, the fact is that the idea of experience is rooted in our educational practices to become one of the top concerns of those seeking to understand or to solve a problem. Universities, institutes and education in general today are based on technical education that prepares students for the expected responsibilities. EBI, which provides education for engineers in applied industrial biology, listed its efforts in this process that commits teachers and students by giving everyone a role to play.

EBI provides a 5 years MSc Diploma course to train students to work as engineers in the field of pharmaceutics, cosmetics, food engineering, environment, and others [1]. In the undergraduate cycle, students learn mathematics, physics, biology and chemistry. Among the courses of physics, "fluid mechanics" [2,3] in 2nd year was considered by students as difficult and they seemed somewhat unmotivated. In order to encourage them to show more interest to this course, a new pedagogy based on atypical experiments was tested on students of year "P17". Students were asked to draw inspiration from everyday life situations to find applications of course and to do atypical experiments to verify and validate some theoretical results.

To assess the relevance of this new pedagogy, we will analyze course evaluations by students of "P17" and we will compare them with those of three other years (two before the new pedagogy and one after). The following study focuses on students of four years: "P15" (2005/2006), "P16" (2006/2007), "P17" (2007/2008) and "P18" (2008/2009).

### 2. Methodology

Our methodology is based on the goals of our institute which aims to provide industry with practical engineers. The education is based on the abilities of the teacher to put into practice the program and give students the opportunity to become more involved. The relationship between teacher and learners determine the learning process to a large part. By empowering the student to enable him to become the cornerstone of the act of learning, the teacher performs half the way of the learning process. Our methodology is therefore based on the idea of motivating students with a play-based pedagogy through atypical experiments to verify and validate some theoretical results found in course by referring to daily life situations.

#### 2.1. Example of atypical experiments in fluid mechanics

As examples of atypical experiments in fluid mechanics made by students of "P17", we present experiments of emptying jerry cans. These experiments of emptying jerry cans were made by three students: Lucie Clavel, Maëla Drouin, and Laure-Anne Gillon. Their work was presented in a report [4] with: Goal of experiments, Material and Method, Experiments and Results, and Conclusions.

2.1.1. Presentation of the experiments

The students began by identifying the required material for these experiments, namely: two plastic jerry cans one blue non-transparent of 35 liters and a second white transparent of 20 liters, two plugs of 1 cm diameter, a stopwatch, a meter, a scales of 100g precision and a spirit level (figure 1). They chose to work with water for a practical purpose: its accessibility and physical properties as density.

They made holes of 1cm diameter in the jerry cans to allow a slow flow, easy to measure with a stopwatch, hence a more accurate measured emptying time. The holes were placed in the bottom part of jerry cans in order to allow an adequate visualization of the flow (figure 1). They were also made to obtain a contraction coefficient $C_c$ equal to 0.61. When discharging, the jerry cans were placed on a coffee table, which horizontality was controlled with a spirit level. Thus, the height of water level measured in the jerry cans is uniform. Before beginning the experiments, the jerry cans filled with water were weighed to determine the mass of water passed by subtracting the mass of the jerry can at the end of emptying.

2.1.2. Experiments and results
By applying the Bernoulli equation from the water surface to the orifice, the flow is obtained as

$$Q = \sqrt{\frac{2gh}{\left(\frac{1}{C_c^2 s^2} - \frac{1}{S^2}\right)}} \quad (1)$$

In this equation, g is the gravity acceleration, h level of water, S the horizontal surface of jerry can, s the surface of hole and $C_c$ the contraction coefficient.
The emptying time is given by [2]

$$T = \frac{2\sqrt{h}}{K} \quad (2)$$

Where: $K = \frac{1}{S}\sqrt{\frac{2g}{\left(\frac{1}{C_c^2 s^2} - \frac{1}{S^2}\right)}}$.

For example for the blue jerry can:

    h = 37 cm
    $C_C$ = 0.61
    s = $\pi r^2$ = 0.785 cm$^2$

$S = l*L = 837$ cm$^2$

And the initial flow $Q = 0.00013$ m$^3$/s $= 0.13$ l/s

They find $K = 0.0025$ m$^{0.5}$ s$^{-1}$ and therefore the theoretical emptying time is $T_{theoretical} = 486.6$ s $= 8$ min 7 s.

The experimental emptying time is $T_{experimental} = 8$ min 40 s.

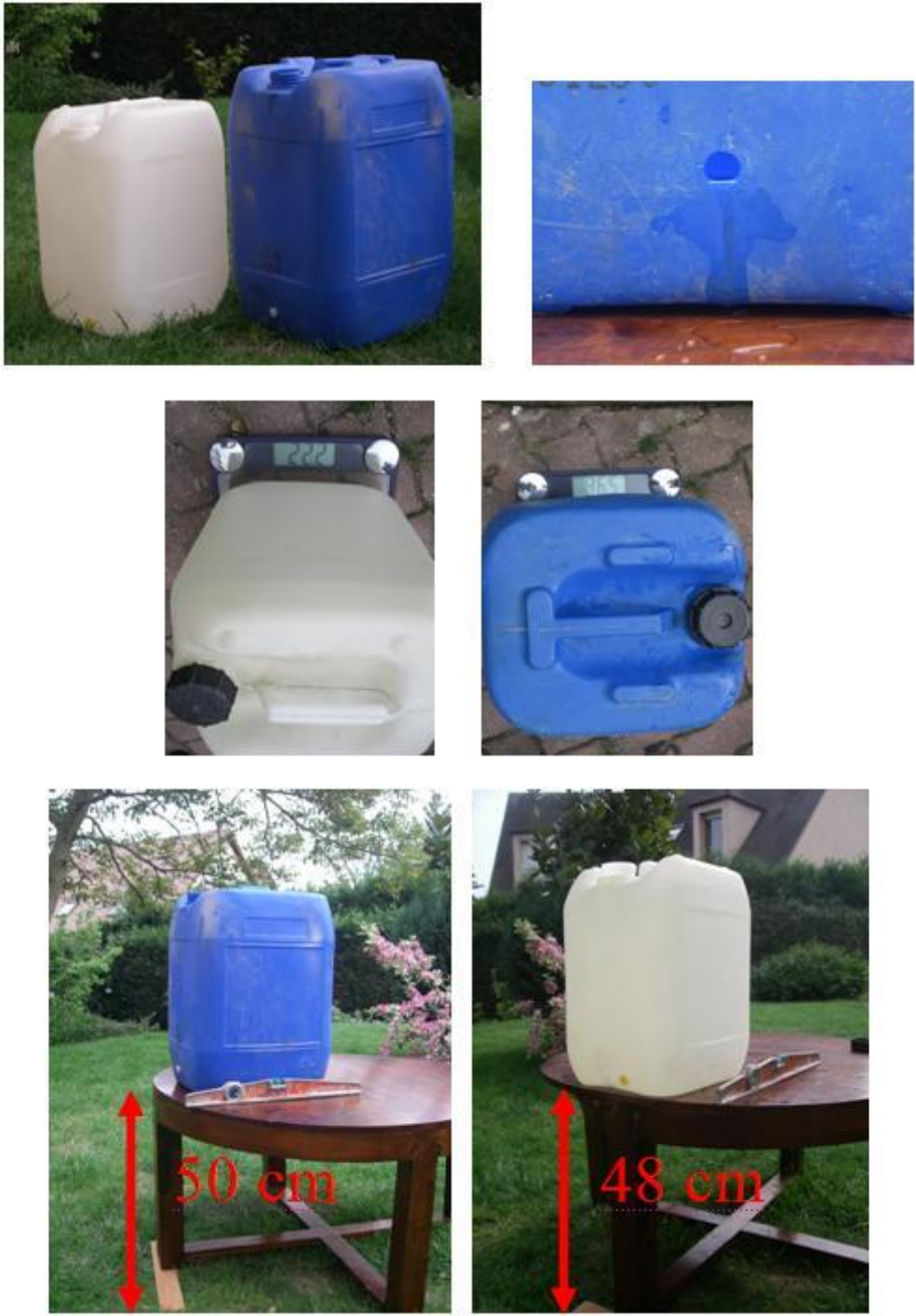

Figure 1: Example of atypical experiments in fluid mechanics: emptying jerry cans, preparation of experiments.

The students found that the theoretical emptying time is lower than the experimental one. Since the result found in course was for a perfect or ideal fluid (µ=0), they explained the difference between theoretical and experimental results by the fact that they neglected the viscosity µ of water in the theoretical time (assumed equal to 0). They concluded that the viscosity of water (µ ≠ 0) will increase the value of the theoretical emptying time and will allow therefore a more accurate value.

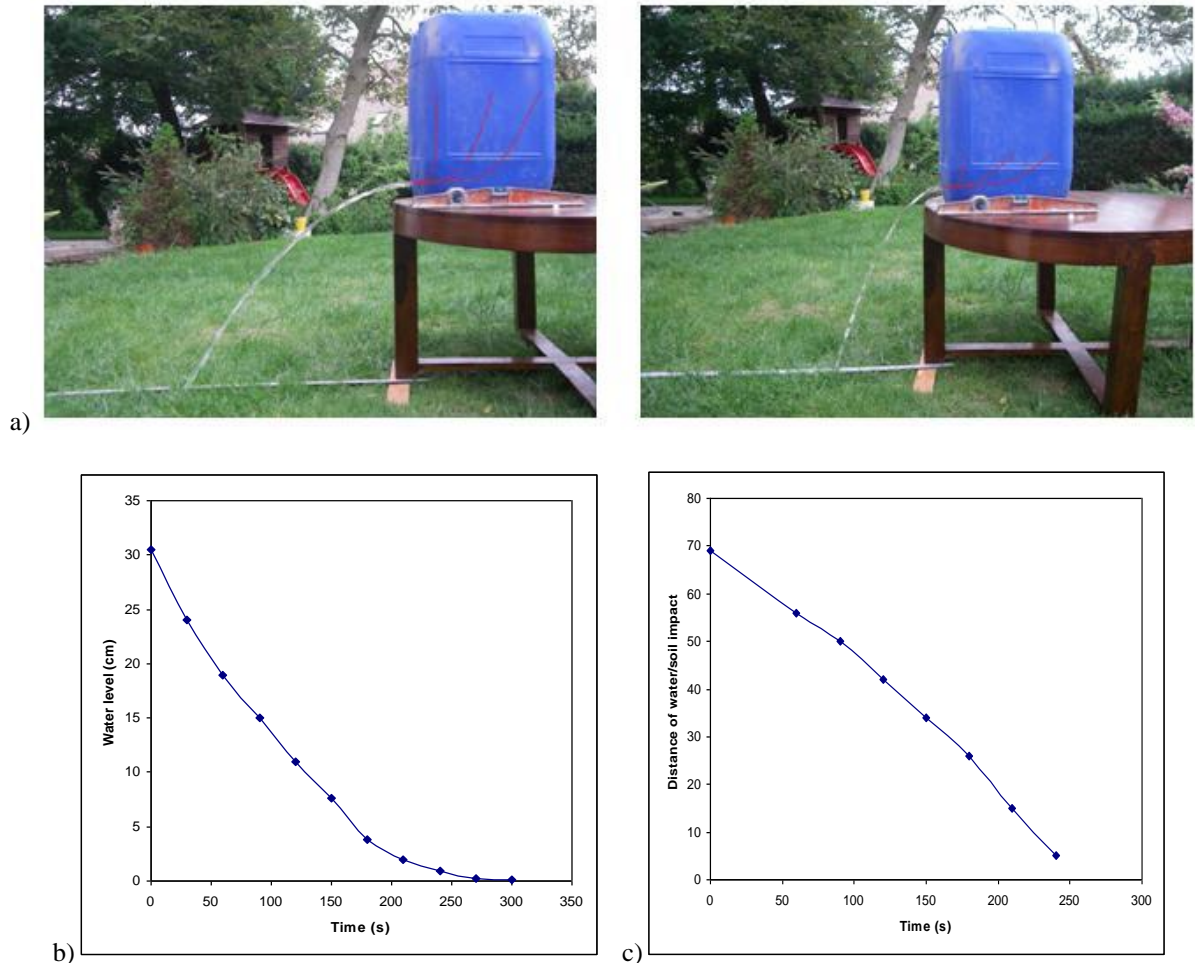

Figure 2: Example of atypical experiments in fluid mechanics: Emptying of a 35 liters plastic jerry can.
(a) Experiments. Evolution of water level (b) and distance of water/soil impact (c) vs time.

### 2.2. Teaching evaluation by students

The impact of our new pedagogy could be assessed from teaching or course evaluation. These evaluations could be considered as a useful tool in order to improve the exchange between teacher and students. Different studies have been conducted on the effectiveness of this tool and its relevance or not [5-8].
Course evaluations by students are performed at EBI on the website of studies.

Process of teaching evaluations by students:
In order to start the online teaching evaluations, students need first to select the course and the year. The evaluation includes two parts (figure 3):
- The first consists on evaluating on a scale of 4: *(1) average, (2) satisfactory, (3) good, (4) very good ;* the following criteria : *organization, required work, clarity of explanations, pedagogy used, interactivity, implication of students, controls.*
- The second consists on answering two questions related to principal forces and issues to be improved. It is also possible to write additional comments and observations.

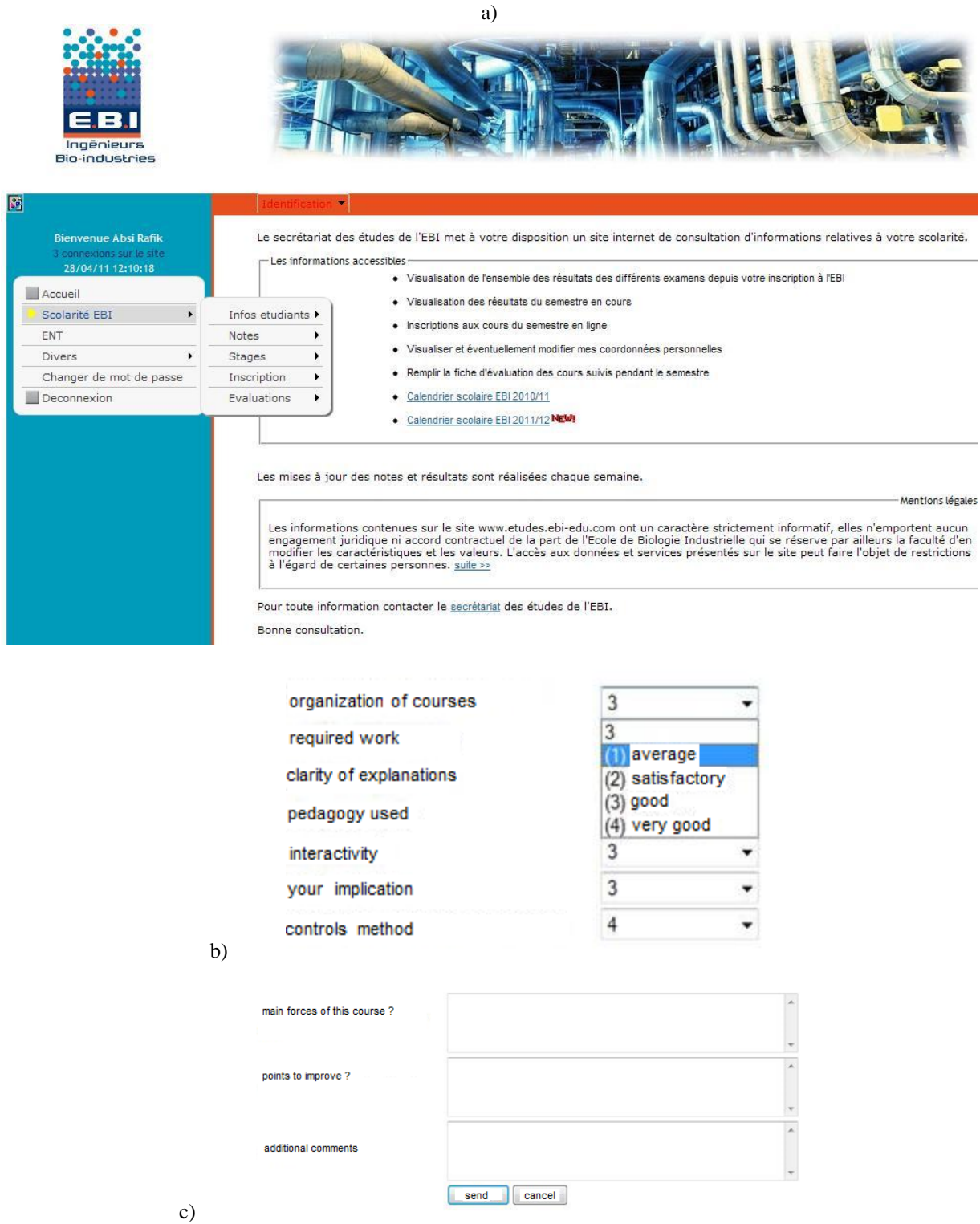

Figure 3: Online course evaluations by students: (a) EBI studies web site, (b) evaluation by "(1) average, (2) satisfactory, (3) good or (4) very good" of each criteria, (c) second part of evaluation: two questions and additional comments.

### 3. Main results and assessment of experience impact on student's learning

#### 3.1. Analysis of teaching evaluation

Figure (4) presents an example of teaching evaluation for fluid mechanics by students of year "P17" [9].

This figure presents the total number of students 122, the number of students who participated in the teaching evaluation 58 and the percentage of participation 47.54%. The different criteria are evaluated on a scale of 4, the average of each criterion is indicated: organization (3.29/4), required work (3.09/4), clarity of explanations (3.47/4), used pedagogy (3.41/4), interactivity (3.28/4), implication of students (2.97/4), controls (3.34/4). The global average of the evaluation is indicated at the bottom 3.26. It shows also three indices respectively: excellence index (40.15%), performance index (87.44%) and satisfaction index (98.77%). The satisfaction index indicates the percentage of students who evaluated all the criteria at least 2/4. The performance index indicates the percentage of students who evaluated all the criteria at least 3/4. The excellence index indicates the percentage of students who evaluated all the criteria 4/4.

| | |
|---|---|
| course | PHY215P- fluid mechanics |
| year | 2008-02-01 session n° 16 |
| teacher | ABSI Rafik |
| total number of student | 122 |
| number of student who participated in the evaluation | 58 |
| percentage of participation | 47.54 |
| excellence index (% of very good) | 40.15 |
| performance index (% of very good + good) | 87.44 |
| satisfaction index (% of very good + good + satifactory) | 98.77 |
| organization | 3.29 |
| required work | 3.09 |
| clarity of explanations | 3.47 |
| used pedagogy | 3.41 |
| interactivity | 3.28 |
| implication of student | 2.97 |
| controls | 3.34 |
| global average | 3.26 |

Figure 4: Sample of teaching evaluation of fluid mechanics course by P17 students.

Figure (5) presents the average of the different criteria for students of four years (P15 to P18). This figure allows observing a peak for students of year P17 who were concerned by the new pedagogy based on atypical experiments. This peak indicates that all criteria have maximum values and therefore that the teaching was considered as the best on the basis of the evaluation of the different criteria namely: organization, required work, clarity of explanations, pedagogy used, interactivity, implication of students and controls.

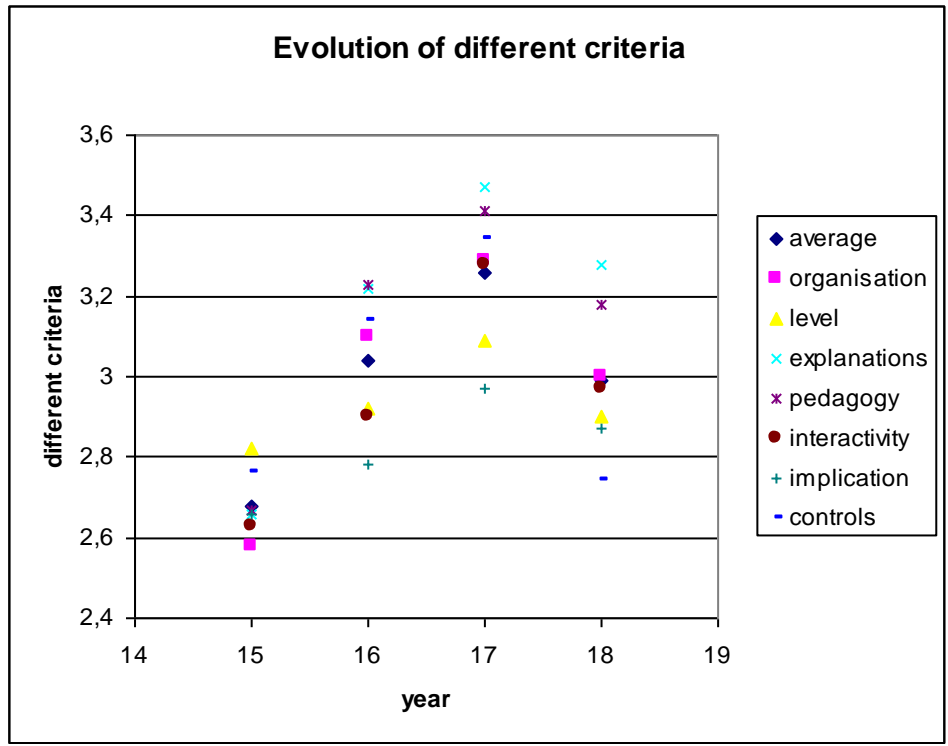

Figure 5: Evolution of different criteria for each year (from P15 to P18)

It is important to note that the number of students who participated in the teaching evaluation has decreased from year P15 to P18 (figure 6). This should have a significant effect on the evaluation results and their analysis.

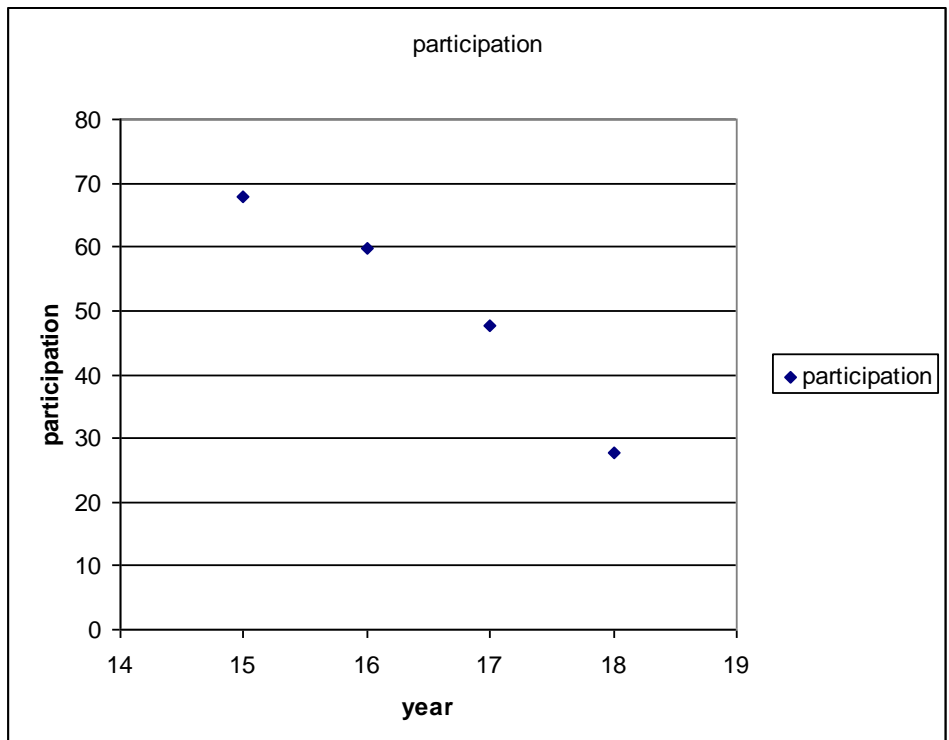

Figure 6: Percentage of students' participation to teaching evaluation (from P15 to P18).

**3.2. Semantic analysis of comments and observations**

The students' comments were analyzed. We identified the different "words" hereafter called "quotations", which were written by students in their comments and observations.
These quotations have been grouped by semantic fields and groups (Table 1).

| Semantic Groups↓ | | Students → Semantic fields ↓ | P18 n | P18 n/N ; N=39 | P17 n | P17 n/N ; N=58 | P16 N | P16 n/N ; N=73 | P15 n | P15 n/N ; N=93 |
|---|---|---|---|---|---|---|---|---|---|---|
| (1) | (+) | Amiability | 6 | 0,1538 | 4 | 0,0689 | 5 | 0,0684 | 3 | 0,0322 |
| | | Passionate | 2 | 0,0512 | 2 | 0,0344 | 4 | 0,0547 | 1 | 0,0107 |
| | | Listening | 2 | 0,0512 | 5 | 0,0862 | 5 | 0,0684 | 4 | 0,0430 |
| | (-) | Authority | 0 | 0 | 1 | 0,0172 | 4 | 0,0547 | 4 | 0,0430 |
| | | Delay | 0 | 0 | 0 | 0 | 0 | 0 | 2 | 0,0215 |
| | | Noise | 0 | 0 | 4 | 0,0689 | 7 | 0,0958 | 3 | 0,0322 |
| (2) | (+) | Experiments, lab work, project | 4 | 0,1025 | 5 | 0,0862 | 3 | 0,0410 | 2 | 0,0215 |
| | | Clarity | 4 | 0,1025 | 9 | 0,1551 | 11 | 0,1506 | 6 | 0,0645 |
| | | Interactivity | 2 | 0,0512 | 7 | 0,1206 | 7 | 0,0958 | 6 | 0,0645 |
| | | Explanations | 5 | 0,1282 | 6 | 0,1034 | 10 | 0,1369 | 9 | 0,0967 |
| | | Demonstrations | 0 | 0 | 2 | 0,0344 | 3 | 0,0410 | 9 | 0,0967 |
| | | Recalls | 0 | 0 | 3 | 0,0517 | 4 | 0,0547 | 6 | 0,0645 |
| | | Motivation | 1 | 0,0256 | 2 | 0,0344 | 4 | 0,0547 | 5 | 0,0537 |
| | | Concrete | 3 | 0,0769 | 3 | 0,0517 | 7 | 0,0958 | 8 | 0,0860 |
| | | PowerPoint | 0 | 0 | 0 | 0 | 0 | 0 | 8 | 0,0860 |
| | (-) | Repetitions | 0 | 0 | 0 | 0 | 1 | 0,0136 | 5 | 0,0537 |
| | | Level difference | 3 | 0,0769 | 0 | 0 | 0 | 0 | 5 | 0,0537 |

Table 1: semantic analysis of students' comments and observations
Semantic groups: (1) Human qualities, (2) Pedagogy; (+) forces, (-) issues to be improved.

Table (1) presents the *occurrence* which is the number of quotations "n" associated to each field by year.
In order to take into account the effect of the number of students who participate in the course evaluation "N", we divided "n" by "N".

Figure (7.a) presents the number of quotations "n" related to each field by year. Figure (7.b) shows the effect of number of responses or number of students who participate in the teaching evaluation "N". The parameter "n/N" for year P17 (solid line) is above the other lines (three other years) for criteria "experience", "clarity" and "interactivity". However, we notice that the criteria "concrete" decreases for P17, while atypical experiments should raise this criterion. We can explain this by the fact that students are not obliged to write comments and therefore some students found enough to rate each specified criterion (on scale of 4). There is some redundancy between the quantitative evaluations related to the specified criteria and free comments.

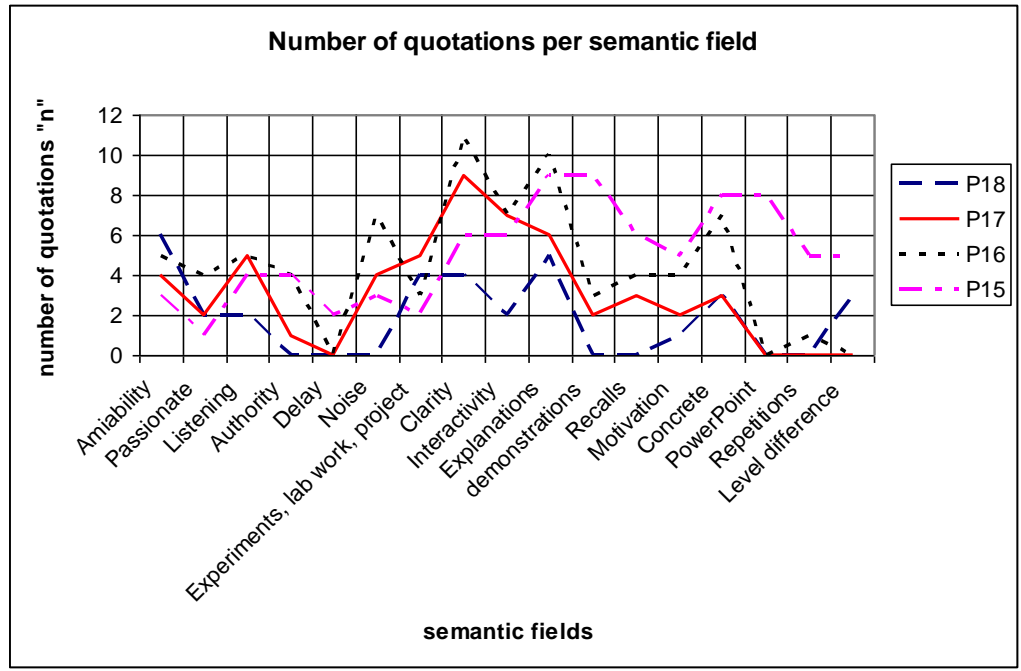

a)

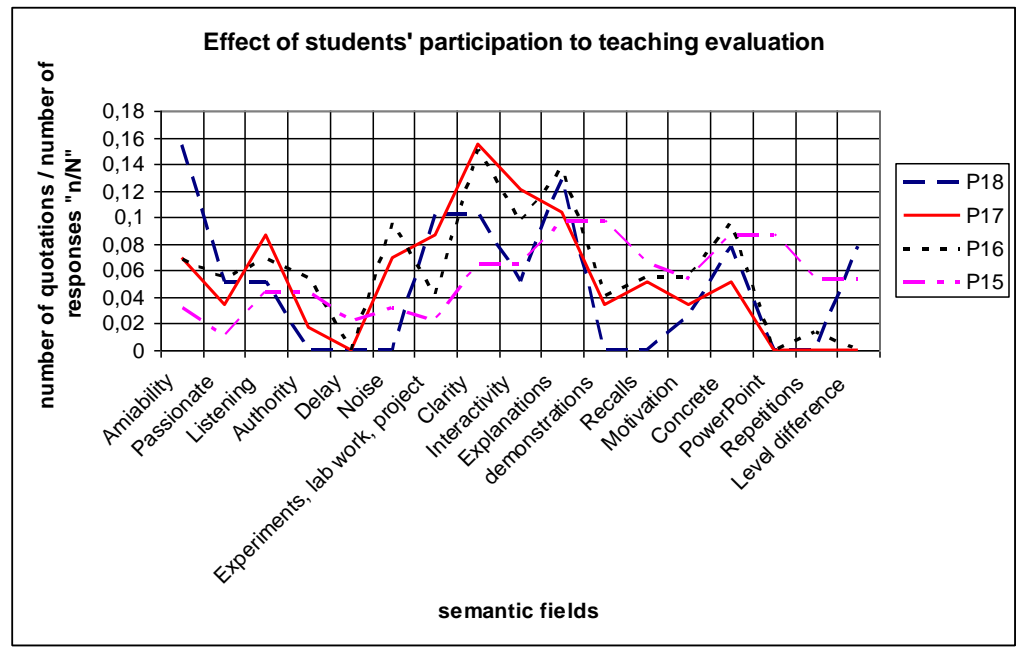

b)

Figure 7: (a) Number of quotations « n » related to each semantic field by year. (b) Effect of the number of students who participated in the teaching evaluation « N », n/N VS semantic fields.

### 3.3. Impact on students' learning

[2]Students' participation in improving teaching through atypical experiments and course assessment gives teaching a new dimension. By this free participation in these experiments, the student discovers the value of effort and no longer hesitates to ask questions or seek solutions to encountered problems. Thus learning is no longer a mere reproduction of abstract knowledge or non-practical applications but an anchorage in the environment.

### 4. Conclusions

In this paper, we presented an innovative teaching/learning pedagogy based on the concept of atypical experiments and its implications in teaching fluid mechanics. This new pedagogy was tested at EBI on students of year "P17". Students of "P17" were asked to draw inspiration from everyday life situations to find applications of course and to do experiments to verify and validate some theoretical results. The experiment of emptying of jerry cans is presented as an example of these atypical experiments. The impact of this innovative pedagogy has been evaluated from online teaching assessment made by students. We analyzed the course evaluations by students of "P17" and we compared them with those of three other years (two before the application of the new pedagogy and one after). The teaching evaluation concerns: (1) a first part which consists on assessing on a scale of 4 the criteria: organization, required work, clarity of explanations, pedagogy used, interactivity, implication of students and controls; (2) a second part which consists on two questions related to main forces and issues to be improved and free comments and observations.

The analysis of teaching assessment shows for part 1 (assessment of criteria on a scale of 4 for years P15 to P18) a peak in all criteria for students of year P17, which was concerned by the application of the new pedagogy based on atypical experiments. This result is very interesting and shows the efficiency of this approach to motivate students. In addition to the quantitative analysis, we analyzed comments and observations of students (part 2 in the teaching assessment). We identified first the main "words" called "quotations" which were grouped into semantic field and groups. We compared first the occurrence which is the number of quotations "n" related to each field for a given year. However, in order to account for "N" the number of students who took part in the teaching assessment, we introduced the parameter "n/N". The parameter "n/N" for year P17 is above the three other years for criteria "experience", "clarity" and "interactivity".

Learning through play proved a great success in fluid mechanics which has been progressively perceived as interesting, useful, pleasant and easy to assimilate. We showed that this pedagogy which includes educational gaming presents benefits for students. These experiments seem therefore to be an effective tool for improving teaching/learning activities in higher education.

We can deduce from this study that trust acquires in the cooperation and learning is based on both knowledge and students involvement. In this exchange, the teacher learns as much as student, because in teaching him to remain curious, he reinforces his skills to seek solutions to encountered problems.

**Rafik Absi** is professor of Fluid mechanics modelling at EBI Cergy-Pontoise since 2002. He received his Diplôme d'Ingénieur from Ecole Nationale Polytechnique of Algiers (Algeria) in 1994 and a Ph.D. in Fluid Mechanics from University of Caen (France) in 2001. He has been a visiting professor at the University of Minnesota, Saint-Anthony Falls Laboratory (USA) and Tohoku University (Japan). His main interests are in analytical modelling, turbulent boundary layers and sediment transport. He is expert and a member of AFSSET (now ANSES, French environment agency) CES Water and Biological Agents.

**Caroline Nalpas** is a graduate student at EBI Cergy-Pontoise.

**Florence Dufour** founded EBI at Cergy-Pontoise, she is Dean of this graduate school and professor of Industrial Quality since 1992. She is veterinary doctor (Alfort 1984) and received her Ph.D. in 1987 from University Pierre et Marie Curie (Paris 6, France) UMPC and INAPG. She is member and expert for CTI (the Engineering Qualifications Commission), AFSSAPS and AERES. Since 2009, she is president of "Commission Ecoles d'Ingénieur et Société" of CDEFI (Conférence des Directeurs des Ecoles d'Ingénieurs Françaises)

**Denis HUET** received a Ph.D. in Enzymatic Engineering, Bioconversion and Microbiology from the University of Technology of Compiègne (France) in 1991. After an industrial experience in the development of biosensors in the context of a join venture, he joined EBI as a professor of immunology and enzymology (1992). In 2005 he became director of studies at EBI. Since, he participates and coordinates the implementation of projects and innovative tools in the field of educational engineering

**Rachid Bennacer** is a professor of universities, he is Mechanical Engineer (1989), and he got his PhD thesis at Pierre et Marie Curie University (Paris 6, France) in 1993. He worked as lecturer in the University Paris XI (1993), become an associate professor at Cergy-Pontoise University 1994. After working as full Professor at Cergy University he moved to the Ecole Normal Superieure (ENS Cachan). His Research fields cover several domains as material science, energy system, pollution and renewable energy with an expertise in convection-diffusion problems. He authored more than 100 international publications, member of several administration or scientist council, organizing committee of conference, journals main board member and invited professor in several international and prestigious universities.

**Tahar Absi** is a professor of universities in psychology and education sciences at University of Algiers. He is Doctor of philosophy in education with a Ph.D. from University of Paris Pantheon-Sorbonne (France) and Doctor in Science Education with a "Doctorat d'Etat" from University of Algiers (Algeria). He was a professor of Philosophy at El Mokrani School. He authored several articles published in scientific journals and newspapers on education, dialogue between cultures and civilizations. He wrote a book on communication and translated books from French to Arabic.